\documentclass[prb,twocolumn,notitlepage,amsfonts,amssymb,amsmath,floatfix,tightenlines,superscriptaddress]{revtex4-1}
\usepackage{bbold}
\usepackage{amsfonts,amssymb}
\usepackage{amsmath,times}
\usepackage{graphicx}
\usepackage[colorlinks=true,linkcolor=blue,anchorcolor=red,citecolor=blue,urlcolor=blue]{hyperref}
\usepackage{color}
\usepackage{cleveref}
\usepackage[lofdepth,lotdepth,caption=false]{subfig}
\usepackage[normalem]{ulem}

\begin{document}

\title{The Higgs-Amplitude mode in the optical conductivity in the presence of a supercurrent: Gauge-invariant formulation with disorder}

\author{Ke Wang}
\affiliation{Department of Physics and James Franck Institute, University of Chicago, Chicago, Illinois 60637, USA}
\affiliation{Kadanoff Center for Theoretical Physics, University of Chicago, Chicago, Illinois 60637, USA}

\author{Rufus Boyack }
\affiliation{Department of Physics and Astronomy, Dartmouth College, Hanover, New Hampshire 03755, USA}

\author{K. Levin}
\affiliation{Department of Physics and James Franck Institute, University of Chicago, Chicago, Illinois 60637, USA}

\begin{abstract}
Observing the ``Higgs" or amplitude mode in superconductors has been a central challenge in condensed matter physics. 
Moreover, arriving at a theoretical understanding of this mode and how it is accessible in, say, conductivity experiments presents an additional challenge as here one needs to satisfy gauge invariance in the presence of disorder. 
In this paper, we characterize the Higgs contribution within a fully gauge-invariant treatment of the linear optical conductivity, $\sigma(\omega)$, for a disordered superconductor carrying a uniform supercurrent.
As a consequence of gauge invariance, there are two distinct charge conservation laws underlying the linear electromagnetic response with two associated sets of $f$-sum rules.
An interesting finding from the Higgs-related sum rule is that the imaginary part of $\sigma(\omega)$ yields an anisotropic, \textit{negative} $1/\omega$ contribution in the THz regime. This is relevant to device applications and appears to be consistent with recent experiments. The work presented here emphasizes how difficult it is to disentangle the neutral amplitude mode contributions from those of the charged quasi-particles and we demonstrate why this is the case.
\end{abstract}

\maketitle

{\it \textbf{Introduction.}} 
There has been a longstanding ambition in condensed matter to observe the superconducting amplitude mode, often known as the ``Higgs" in the language of particle physics~\cite{Higgs,Guralnik1964,Englert1964, GHKBook, footnote}.
It has been recognized
~\cite{Volovik2014,Pekker2015,Shimano2019} for some time that there is a way to observe this mode even in the linear electromagnetic
response, beginning with the work of Littlewood and Varma~\cite{Littlewood} and our own group~\cite{LevinBrowne} (Levin and Browne) 
who considered particular kinds of superconductors containing a co-existent charge-density wave.
More recently, this mode has been predicted~\cite{Efetov2017} and observed to be present~\cite{Shimano2019} in superconductors that carry a time-independent or static supercurrent. It has similarly been observed
in non-linear applications (involving time-dependent electromagnetic potentials) corresponding to a class of third harmonic generation experiments~\cite{Cea2016,Seibold2021,Silaev2019,Tsuji2020,Murotani2019}.

This paper deals with the problem of arriving at a proper theoretical understanding of this Higgs mode in the linear AC conductivity when such a static supercurrent is present. 
We emphasize the challenge, as one has to first prove that gauge invariance is present in the theory and that the amplitude mode is necessarily incorporated into this proof. 
Secondly, one has to establish gauge invariance in the presence of substantial disorder, as we consider conductivity experiments which necessarily incorporate this disorder.
We note that there are earlier schemes~\cite{Efetov2017,Kubo} that utilize the Eilenberger-Usadel approach, and although these represent very nice contributions, they are appropriate to the case of extreme disorder and are thought to suffer from inconsistencies with gauge invariance~\cite{Kita,Gorini2010,Yang2022}. 

While earlier work has advanced our understanding of the dynamics of the Higgs mode~\cite{Cea2016,Seibold2021,Silaev2019,Tsuji2020,Murotani2019,Yang2019,Matteo2020,Lee,Moore2022,Liang2022,Lee}, a fully gauge-invariant electromagnetic (EM) response theory that incorporates all collective modes and accommodates arbitrary disorder in the presence of a supercurrent is not yet available.
Our paper establishes such a response theory. Gauge invariance ensures the collective modes can be grouped into two sectors: a \emph{charge channel}, which consists of phase fluctuations and quasiparticle excitations, and a \emph{neutral Higgs channel}. Each produces a gauge-invariant optical response, leading to two distinct $f$-sum rules. 
We find that the Higgs channel is associated with a negative superfluid density contribution to the anisotropic conductivity, leading to a $-1/\omega$ dependence in the anisotropic imaginary component within the THz regime. This prediction is supported by experimental indications~\cite{privatecommunication} and is expected to be relevant to device applications~\cite{Kubo,Jonas2012}.
We show how and why the Higgs channel is difficult to disentangle in the optical response from the charge channel. We find, moreover, that the strength of the disorder governs the size of the relative contributions of the two channels. 
Finally, we clarify here that the anisotropic response in the supercurrent scenario is not solely associated with the Higgs contribution, as has been claimed~\cite{Efetov2017} to be the case, except in the limit of very extreme disorder.

\vskip1mm
{\it \textbf{Gauge-invariant linear response.}} In a gauge-invariant formalism one has to incorporate all possible couplings of the current to the vector potential at the linear-response level. The approach we use here was developed by Kulik \textit{et al}~\cite{Kulik1981}.
In contrast to a uniform superconductor, where the Higgs mode is undamped at zero wave vector and independent of impurity effects,  in a non-uniformly paired superconductor or in the presence of a supercurrent where there is broken time-reversal symmetry, the Higgs mode is sensitive to inevitable non-magnetic impurities~\cite{Maki1963,Maki1963II,nosov2024}.
While this greatly increases the complexity of the problem, it is of sufficient importance (with implications for experimental devices~\cite{Kubo}),
that it merits close examination.

  \begin{figure*}
    \includegraphics[width=5.0in,clip]
{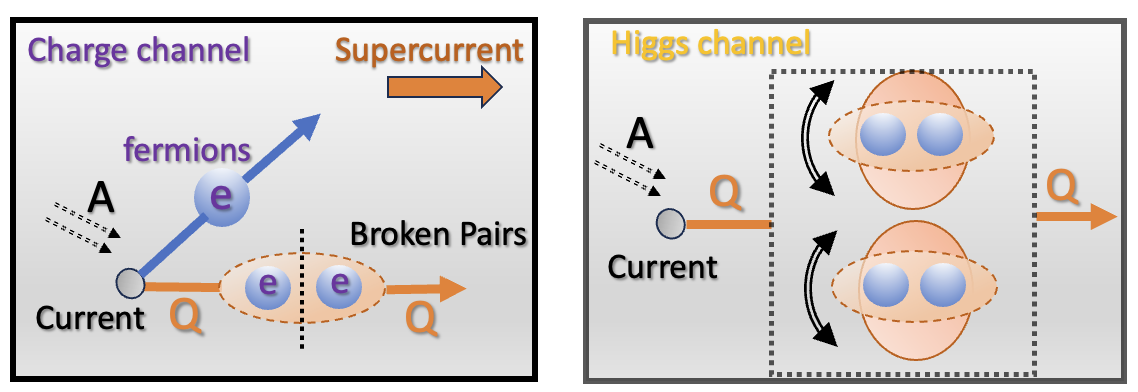}
\caption{Two contributions to the conductivity when a supercurrent is present. Charge channel on the left: The dashed arrows represent the incident electromagnetic (EM) field. The current has two contributions: one component (blue line) originates from fermions, as described in the Mattis-Bardeen conductivity, while the other component (orange line) arises from pair-breaking processes induced by the supercurrent and disorder. 
Higgs channel on the right (see Eq.~\eqref{eq:Higgs}). 
${\bf Q}$ enables a coupling between the neutral Higgs mode and the current. Black solid arrows depict the collective motion of the order parameter (Higgs mode). Consequently, the charge channel contribution incorporates the effects of broken pairs, leading to behavior that closely resembles contributions from the Higgs channel, as illustrated in subsequent figures.} 
    \label{fig1}
\end{figure*}

\vskip1mm
Our work is based on a linear-response formalism~\cite{Kulik1981}
that includes the phase and amplitude fluctuations of the order parameter. Fundamental to linear-response theory is the EM kernel \( K^{\mu\nu} \), which describes how the four-vector
current \( J^{\mu} = (\rho, \mathbf{J}) \) responds to a weakly applied external EM field \( A^{\mu} = (\phi, \mathbf{A}) \):
\begin{equation}
J^{\mu}(q) = K^{\mu\nu}(q) A_{\nu}(q).
\end{equation}
Here $q = (q_0, \mathbf{q})$  is a four-vector momentum. The invariance of the current under the gauge transformation \( A_\mu \rightarrow A_\mu + \partial_\mu \lambda \) necessitates that \( q_{\mu} K^{\mu\nu} = 0 \), which is a crucial condition for a consistent linear-response theory. 

One can further decompose the kernel into 
\begin{equation}
  \label{eq:em4}
  K^{\mu\nu}(q) = K_0^{\mu\nu}(q) + {K}_1^{\mu\nu}(q)\;.
\end{equation}
Here, $K_{1}$ accounts for the collective amplitude and phase motion of the condensate of Cooper pairs. $K_{0}$ reflects the fermionic channel, where
\begin{equation}
K_0^{\mu\nu}(q) = P^{\mu\nu}(q) +
\frac{ne^2}{m}G^{\mu\nu}.
\label{3}
\end{equation}
The retarded current-current correlation function is defined in real space as $P^{\mu\nu}(t,{\bf x})\equiv
-i\theta(t)\langle[\hat j^{\mu}(t,{\bf x}),\hat j^{\nu}(0,0)]\rangle$ where $\hat j$ is the current operator. 
Here $G^{\mu,\nu}$ is a $4 \times 4$ diagonal tensor whose diagonal elements are $(0, 1, 1, 1)$, representing a diamagnetic contribution.

To derive the $K_1$ response, we implement a matrix linear-response approach \cite{Kulik1981} in which the perturbation of the condensate $\Delta_1+i\Delta_2$ is included as additional contributions to
the perturbing external field. These external fields $(A_\mu, \Delta_1, \Delta_2 )$ cause a non-zero  current $\langle j_\mu \rangle$ and order parameters $-g\langle \hat\eta_{1/2} \rangle/2$. Here 
$\hat\eta_{i} $ is the superconductor pairing field defined by
$ \Psi^\dagger(t,{\bf x})\tau_i \Psi(t,{\bf x})$, where $\Psi\equiv ( \psi_\uparrow,  \psi^\dagger_\downarrow)^T$, $\psi_{\uparrow/\downarrow}$ is the spin-up/down fermionic operator, $\tau_i$ is the $i$-th Pauli matrix and $g$ is the attractive $s$-wave interaction strength. For convenience, we let $\mu=0,1,2,3$, and $i=1,2$. Eliminating $\Delta_{1,2}$ leads to~\cite{Levin1995,Rufus2020} 
\begin{equation}
  {K}^{\mu \nu}_1(q)=  -\sum_{i,j=1}^2 R ^{\mu i}(q) \left(S+\frac{2}{g}  \right)_{ij}^{-1} R_t^{j \nu}(q).
\label{eq:4}
\end{equation}
Here $R$ and $R_t$ are the cross correlation function between the current operator and the superconductor pairing field, $2/g$ 
relates to the interaction strength, and $S$ is a correlation function involving the two superconductor pairing fields. The real-space definitions are given by
$R^{\mu i}(t,{\bf x})=-i\theta(t)\langle[j^{\mu}(t,{\bf x}),
\hat{\eta}_i(0,0)]\rangle$, $R^{i\mu}_t(t,{\bf x})=R^{\mu i}(-t,-{\bf x})$, and
$S_{ij}(t,{\bf x})=-i\theta(t)\langle[\hat{\eta}_i(t,{\bf x}),
\hat{\eta}_j(0,0)]\rangle$, with $i,j=1,2$. 
\footnote{The formalism can be easily generalized to include Coulomb fluctuations. Since we are mainly concerned with frequency comparable to $\Delta$, we focus on order-parameter fluctuations.}

For all but strongly interacting superconductors, we can presume particle-hole symmetry around the Fermi surface so that the mixing term between phase and amplitude vanishes (i.e., $S_{12}=0$). In this scenario, contributions from phase and amplitude fluctuations of order parameters can be separately written as
\begin{align}
\label{5}
K^{\mu \nu}_{\text{phase}}&=-\frac{R^{\mu
2}R_t^{2\nu}}
{S_{22}+2/g } ,\\
K^{\mu \nu}_{\text{Higgs}}&=- \frac{R^{\mu
1}R_t^{1\nu}}
{S_{11}+2/g }. \label{eq:Higgs}
\end{align}
It follows that one can construct two separate gauge-invariant components for the response tensor: 
$K_{\text{charge}}\equiv ~K_0+K_{\text{phase}}$ and $ K_{\text{Higgs}}$, where $K_{\text{charge}}$ includes fermion dynamics and phase fluctuations. 

We now prove gauge invariance of the two response tensors in a clean system, and then later we will address the
more complicated proof for the dirty case. We start from the Ward-Takahashi identity (WTI) representing $U(1)$ symmetry in Nambu space~\cite{Guo2013,Rufus2016,watanabe2024}:
\begin{eqnarray}
\label{Ward}
\tau_3 G_0^{-1}(p_+)-G_0^{-1}(p_-)\tau_3 =q_\mu \gamma^\mu+2i\Delta\tau_2.
\end{eqnarray}
Here $G_0$ is the fermionic propagator, $p_\pm=p\pm q/2$, $\gamma^\mu=(\tau_z, {\bf p}\tau_0)$ and $\Delta$ is the order parameter. One can use the WTI to prove the following identities:
\begin{align}
q_\mu K_0^{\mu\nu}(q)&=-2i\Delta R^{2\nu}, \label{7} \\
q_\mu R^{\mu i}(q)&= -2i\Delta \left(S_{2i}+\frac{2}{g} \delta_{2i} \right) \label{8} .
\end{align}
With these equations, it follows that
\begin{equation}
\label{9}
q_\mu K^{\mu \nu}_{\text{charge}}=0,\quad q_\mu K^{\mu \nu}_{\text{Higgs}}=0.
\end{equation} 

The fact that there are two separate gauge-invariant response tensors is particularly important. This reflects the presence of the Higgs mode and the statement that there are two separately conserved charges. 
It should be emphasized that it is easy to miss~\cite{Moore2022,Liang2022,Lee} the amplitude mode, as it is separately gauge invariant. Rather, when checking gauge invariance of a theory, there is not just one but two constraints to be verified through Eqs.~\eqref{7} and \eqref{8}. This is a central point of this paper.
Moreover, since the Higgs mode is a neutral excitation, the charge involved in $K_{\text{charge}}$ is the total charge of the system while the charge in the Higgs channel is zero. What is also most important and will be challenging for any theory is to show, as we do here, that the analogue equations are also satisfied in the presence of impurities.

Turning to a more physical understanding, in Fig.~\ref{fig1}, we illustrate the way these two terms $K_{\text{charge}}$ and $K_{\text{Higgs}}$ contribute to the conductivity and how they appear to have similar frequency structure. 
This observation depends on pronounced disorder effects, as will be discussed in more detail later.
As shown in the figure on the left, in the presence of the supercurrent and disorder, the charge channel contains a new term which depends on the superconducting amplitude. Nevertheless, this is distinct from the Higgs mode, as it is associated with charge effects and more particularly with the contribution to the current from impurity-derived broken Cooper pairs.
In contrast, the figure on the right characterizes the way in which the neutral Higgs mode appears in the electromagnetic response, via coupling of the form $\mathbf{A} \cdot \mathbf{Q}$.

 \begin{figure}[t] 
\includegraphics[width=3.0in,clip]{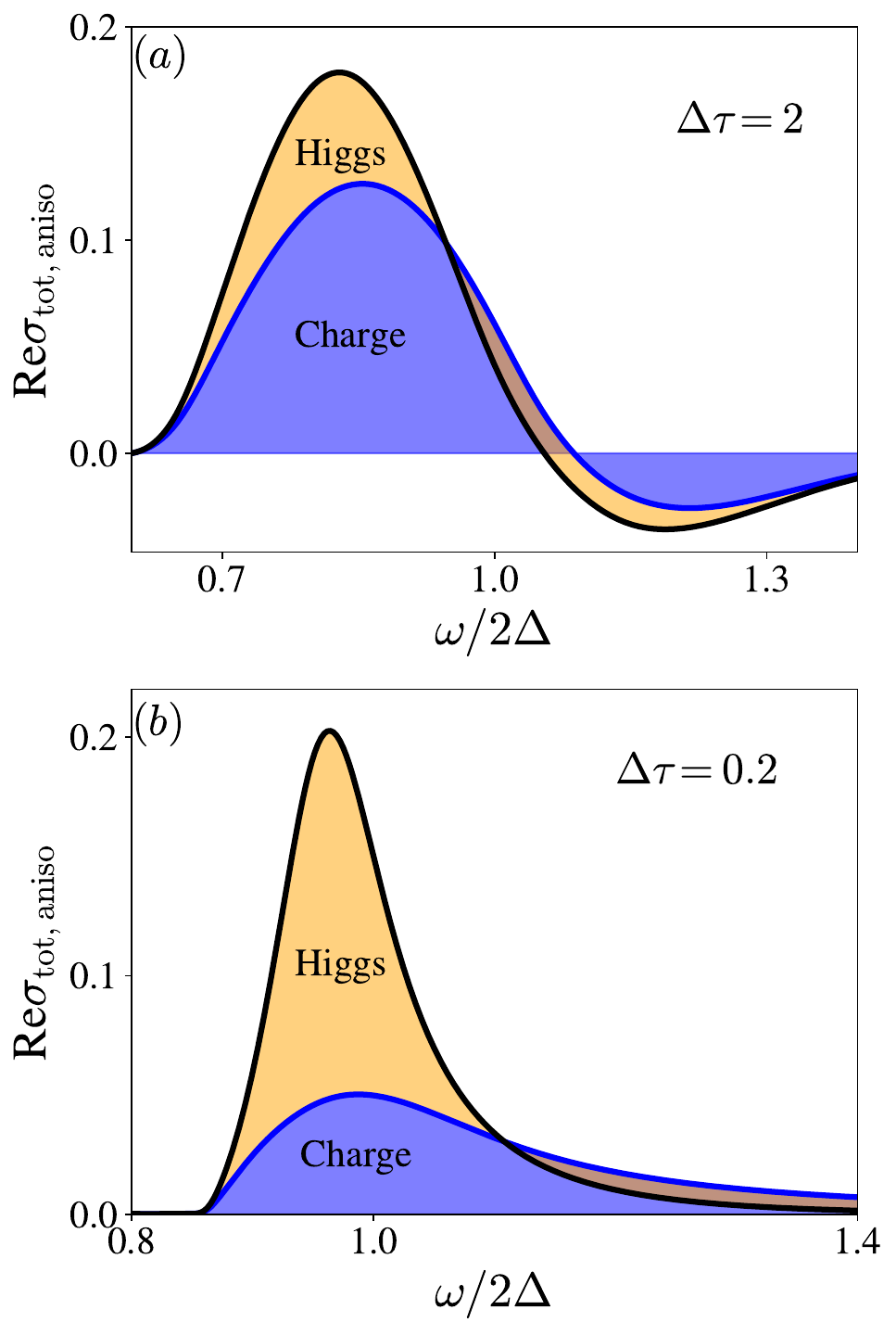}
    \caption{
Plot of the decomposition of the \textit{total anisotropic} component of the conductivity given by
$\sigma_{\text{tot},\text{aniso}} = 
\hat{Q}_i \sigma^{ij}_{\text{tot}} \hat{Q}_j - \hat{Q}^\perp_i \sigma^{ij}_{\text{tot}} \hat{Q}^\perp_j$ in units of $\sigma_0= e^2 m k_F / (2\pi^2 \hbar)$, where $\hat{Q}^\perp$ is an arbitrary unit vector perpendicular to $\hat{Q}$. The supercurrent momentum is chosen as $k_FQ/m=\Delta/2$.
(a) Contrary to the claim in Ref.~\onlinecite{Efetov2017} that only the Higgs component is anisotropic, we find that the charge component (which includes quasi-particle currents and phase fluctuations) also exhibits anisotropy and can even dominate over the Higgs component in cleaner samples ($\Delta \tau > 1$). In this regime, both the 
anisotropic charge and Higgs components display a negative tail at $\omega > 2\Delta$.
(b) At the boundary separating clean and dirty samples ($\Delta \tau = 1$), the contributions from the Higgs and charge components become comparable.
(c) As the system becomes dirtier ($\Delta \tau < 1$), the Higgs component dominates the peak of 
the anisotropic conductivity near $2\Delta$. For
$\omega > 2\Delta$
there is a tail in the anisotropic
conductivity which primarily originates from the charge component.}
\label{fig2}
\end{figure}

\vskip2mm
{\it \textbf{Summary of central results}.}
It is useful at this point to summarize some of the central results of this paper which
are related to the mechanisms in Fig.~\ref{fig1}, before
presenting all theoretical details. 
In Fig.~\ref{fig2}, we show the anisotropic contribution to conductivity, decomposed into charge and Higgs sectors. We plot the difference between the parallel and perpendicular components of the total conductivity, represented by \(\hat{Q}_i \sigma^{ij}_{\text{tot}} \hat{Q}_j - \hat{Q}^\perp_i \sigma^{ij}_{\text{tot}} \hat{Q}^\perp_j\), with \(\hat{Q}^\perp\) as a unit vector perpendicular to \(\hat{Q}\). We argue that plotting this difference better highlights the Higgs contribution than comparing the total conductivity with and without a supercurrent~\cite{Shimano2019}. The latter case involves a conventional Mattis-Bardeen~\cite{Mattis1958} contribution, but we emphasize that this does not accurately represent the isotropic component \(\sigma_{\text{iso}}\), as this also depends on the supercurrent.

This intimate connection between the charge and Higgs contributions is evident in these figures, as they can be seen to have most of their structure in a similar frequency range. This makes it extremely difficult to disentangle the Higgs from the background charge contribution.
We also see that the Higgs contribution can be larger or smaller than this anisotropic charge component, depending on the degree of disorder. The relative size of the two contributions reflects their distinct dependence on disorder (see Ref.~\onlinecite{supp}).
One should also note that the Higgs peak does not precisely align with $2\Delta$.
It is also affected by the excitation gap $\Delta_{\text{ex}}$, which is distinct from the order parameter. 
Finally, we note that the anisotropic charge component was assumed to be zero in earlier literature~\cite{Efetov2017}, whereas here we find that it plays a major role in linear response. However, with the inclusion of extremely strong disorder, it will become progressively smaller than the Higgs contribution to the real part of the anisotropic conductivity. Importantly, however, this charge component
can be quite significant when it comes to the imaginary part of the
anisotropic conductivity, as we will see below.

\vskip1mm

{\it \textbf{Two $f$-sum rules.}} 
 We now turn to calculations of the conductivity tensor which lead to these results and depend on
the response tensor:
\begin{equation}
\sigma^{ij}(\omega) = \frac{i e^2}{\omega} K^{ij} (   \omega,\mathbf{q} = 0).
\end{equation}
Here $1\leq i,j\leq 3$ denote three Cartesian directions. One can construct two gauge-invariant contributions to the conductivity $\sigma_{\text{charge}}$ and $\sigma_{\text{Higgs}}$ corresponding to $K_{\text{charge}}$ and $K_{\text{Higgs}}$. As a consequence of gauge invariance, these are importantly associated with separate sum rules. In the presence of a supercurrent, we write the conductivity as a sum of two terms
involving the ``charge" contribution which involves the currents and the phase degrees of freedom, and the Higgs which involves the amplitude fluctuation:
\begin{equation}
\sigma_{\text{tot}}(\omega) = \sigma_{\text{charge}}(\omega) + \sigma_{\text{Higgs}}(\omega).
\end{equation}
We further decompose this ``charge" contribution into isotropic (iso) and anisotropic (aniso)
components and make the same decomposition for the superfluid densities associated with the
charge component, which are of the form 
$n_{\text{iso}}$ and 
$n_{\text{aniso}}$. Below, we adopt the units $\hbar=k_B=m=e=1$. 

Thus, we can write the conductivity associated with the charge sector as :
\begin{align}
\text{Re}\sigma^{ij}_{\text{charge}}(\omega)
&=\pi  \delta (\omega)\left( \delta_{ij} n_{\text{iso}} - \hat{Q}_i |n_{\text{aniso}}| \hat{Q}_j \right) \nonumber \\
& \quad+ \left(\delta_{ij} \sigma_{\text{iso}} + \hat{Q}_i \sigma_{\text{aniso}} \hat{Q}_j \right).
\label{12}
\end{align}
Here, $\hat{Q} = {\bf Q} / |{\bf Q}|$ is a unit vector. The first term, $\propto \delta(\omega)$, represents 
the contribution from the condensate electrons, while the second term describes the conductivity at finite frequencies, as measured in experiments. The $f$-sum rule for the charge component leads to:
\begin{eqnarray}
\int \frac{d\omega}{\pi} \text{Re}\sigma_\text{iso} = n - n_\text{iso}; \quad 
\int \frac{d\omega}{\pi} \text{Re}\sigma_\text{aniso} = +|n_\text{aniso}|.
\label{eq:13}
\end{eqnarray}
Note that the anisotropic superfluid density $n_\text{aniso}$ enters as a negative weight~\cite{Rufus2017}.

For the Higgs mode, which has only anisotropic contributions, we have
\begin{equation}
\label{14}
\text{Re}{\sigma}_{\text{Higgs}}^{ij}(\omega)
= -\pi  \delta (\omega) \hat{Q}_i |n_{\text{Higgs}}| \hat{Q}_j
+  \hat{Q}_i \sigma_{\text{Higgs}} \hat{Q}_j.
\end{equation}
The $f$-sum rule associated with the Higgs contribution is given by
$\int  {d\omega} \text{Re}\sigma_{\text{Higgs}} / \pi = |n_{\text{Higgs}}|$.
As with the anistropic contribution to the charge contribution, the presence of the Higgs depresses the superfluid density and enters as a negative superfluid weight.

\begin{figure}[h] 
\includegraphics[width=3.0in,clip]
{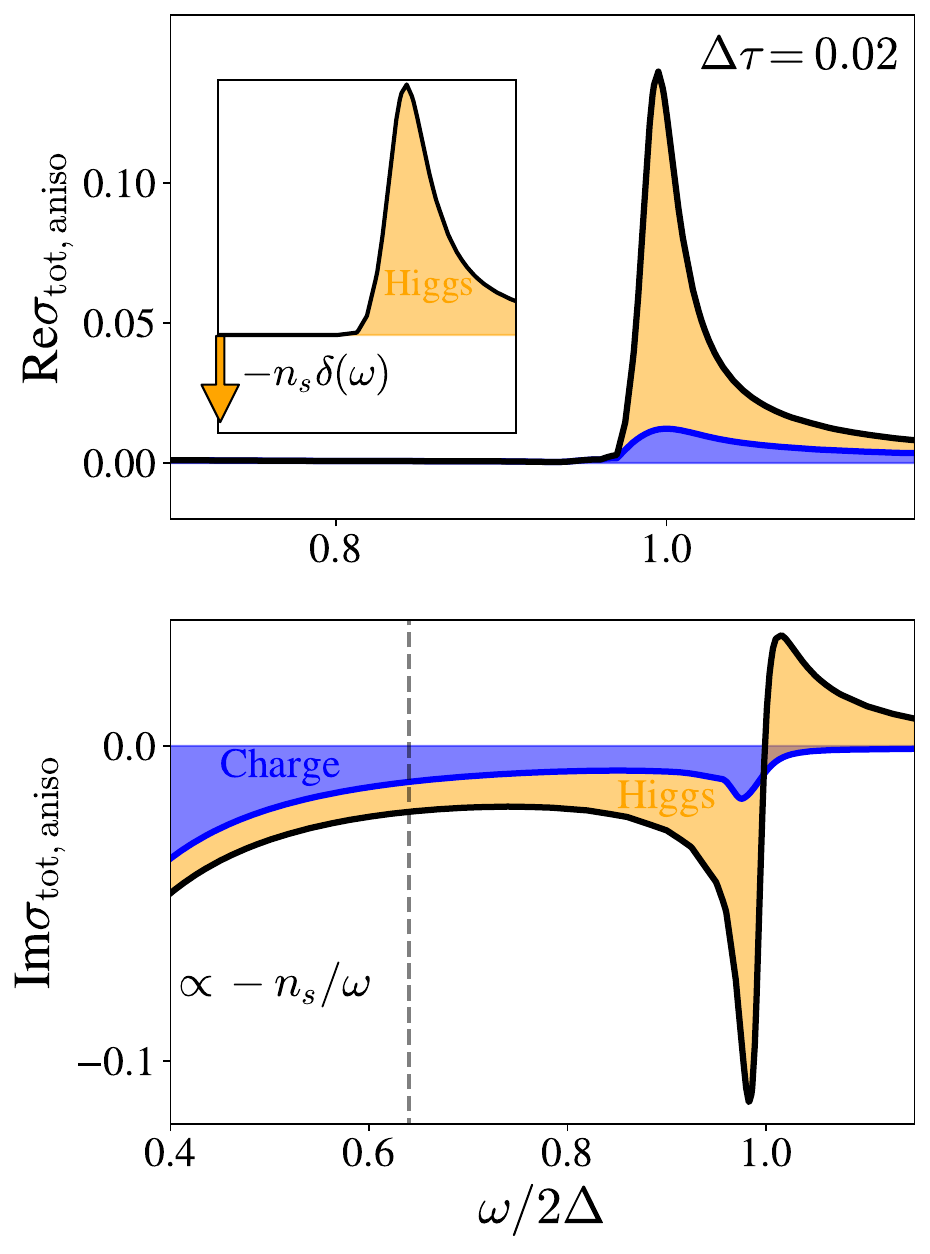}
    \caption{ 
(Color online). The effects of increasing disorder.
Plots of the real and imaginary components of $\sigma_{\text{tot},\text{aniso}}$
in units of $\sigma_0$.  
In the imaginary contribution, an asymmetric feature appears around $\omega = 2\Delta$
and at even lower $\omega < \Delta$, a negative asymptote proportional to $-1/\omega$ emerges, reflecting a negative superfluid weight
in this anisotropic Higgs component, as shown in the inset.
}
\label{fig3}
\end{figure}

\vskip1mm

{\it \textbf{Effects of non-magnetic disorder.}}
In the presence of a supercurrent which breaks time-reversal
symmetry, all disorder leads to pairbreaking ~\cite{Maki1963,Maki1963II}. This will affect
the self consistent conditions for
the central parameters in the theory (the current and the gap) as well as the electrodynamics.

The mean-field Hamiltonian in the presence of a supercurrent can be written in terms of Pauli matrices as
\begin{equation}
H({\bf p})=  {\bf p}\cdot {\bf Q}   +\xi({\bf p}) \tau_z+ \Delta \tau_x.
\end{equation}
Here, $\xi$ represents the fermionic dispersion which we take to be quadratic.
We consider non-magnetic impurities which enter through the term $V({\bf r})\tau_z$
and presume a contact interaction for this disorder.
The self energy, evaluated in the Born approximation, is given by
$\Sigma(p)=(2\pi)^{-3}\int  d^3 {\bf k} V({\bf k})G(p-k)V(-{\bf k})$.  The (inverse) fermion Green's function is then given by $G^{-1}(p)=G^{-1}_0(p)-\Sigma(p)$ with
$G_0(i\omega_0,{\bf p})=[i\omega_0-H({\bf p})]^{-1}$.
Once the Green's function is known, one can determine quantities such as the order parameter $\Delta$, the current $J$, and the single particle excitation gap $\Delta_{\text{ex}}$, which is distinct from the order parameter. 

Disorder has a profound effect on the EM response, which enters through a
Bethe-Salpeter equation for the EM vertex:
\begin{eqnarray}
\label{BS}
\Gamma(p_+,p_-)=\gamma(p_+,p_-)+
\int \frac{d^3k}{(2\pi)^3} V(k) \tau_z
G(p_+')  \nonumber \\
\times \Gamma (p_+',p_-') G(p_-')  \tau_z V(-k),\quad p'_\pm=p_\pm +k.
\end{eqnarray}
Here, $\gamma(p_+,p_-)$ represents a bare (vertex) operator  while $\Gamma(p_+,p_-)$ is its renormalized version. Here, we denote the renormalized version of \(\tau_i\) by \(\Gamma_i\) (with \(i = 1, 2\)) and the current operator $j_\mu$ by \(\Gamma_{\text{J},\mu}\) (with \(\mu = 0,1,2,3\)). For details see Ref.~\onlinecite{supp}.
Then, with all the impurity-renormalized vertices known, one can derive the full set of correlation functions $Q$, $R$, $R_t$, and $S$. For example, the current-current correlation function is 
\begin{eqnarray}
Q_0^{\mu\nu}(k) &=&T\sum_{ip_0}\int  \frac{d^3p}{(2\pi)^3} j_\mu
 G(p)  \Gamma_{\text{J},\nu}
 G(p-k).
\end{eqnarray}
The evaluation of $R$, $R_t$, and $S$ proceeds in a similar way, where either one of two vertex operators is replaced by its renormalized version.

We now come to one of the most important consequences of this section. We are able to prove the Ward-Takahashi identity in the presence of disorder. It can be shown that Eq.~\eqref{Ward} generalizes to:
\begin{eqnarray}
\tau_3G^{-1}(p_+) - G^{-1}(p_-)\tau_3 = q^\mu \Gamma_{\text{J},\mu}  - 2i \Gamma_2 \Delta.
\end{eqnarray}
Using this identity, one can verify that Eqs.~\eqref{7}-\eqref{8} remain valid, and consequently, so does Eq.~\eqref{9}. Thus, we have derived two separately gauge-invariant response tensors for these disordered superconductors in the presence of a supercurrent.

\vskip1mm 
{\it  \textbf{Im$[\sigma]$ in the limit of extreme disorder}}. The
imaginary part of the conductivity contains important information about the superfluid density.
This is plotted in Fig.~\ref{3} along with the real counterpart
for a more highly disordered system. With increasing disorder, we see that the Higgs contribution becomes more dominant. 
There is also a sharpening of the Higgs features in this dirty limit, as impurity effects tend to wash out the anisotropy in $\mathbf{Q}$, which is responsible for damping and pairbreaking, thereby leading to a restoration of Anderson's theorem~\cite{Anderson2}.

Particularly interesting is the negative $1/\omega$ contribution
or downturn with decreasing $\omega$ in the imaginary part of the anisotropic conductivity, as is shown in the inset in Fig.~\ref{3}. 
Importantly, this effect is seen even in the THz regime and is required by consistency with the amplitude-mode sum rule. 
Indeed, this negative sign is different from theory plots in Refs.~\onlinecite{Shimano2019} and \onlinecite{Kubo} where the former
was based on
Ref.~\onlinecite{Efetov2017}. 
Although it was not discussed in any detail, 
experimental evidence for this downturn can be found in Ref.~\onlinecite{Shimano2019} in their Methods section.
 
\vskip1mm

{\it \textbf{Conclusions.}}
The central goal of this paper has been to observe signatures of the Higgs through the linear response conductivity.
Understanding this conductivity requires treating impurity effects, as without disorder an important consequence of Galilean invariance is a vanishing AC conductivity.
It should also be clear that disorder amplifies the relative importance of the Higgs by reducing
the size of the (background) contributions in the charge channel. For this reason this amplitude mode is more
readily observable in very dirty superconductors.
We stress that a combined treatment of gauge invariance and disorder (having variable
strength) is extremely difficult and although there
is a large literature~\cite{Seibold2021,Yang2019,Silaev2019,Tsuji2020,Murotani2019,Cea2016}
particularly on non-linear response, this literature has focused on one or the
other but not both. Addressing both is an important contribution of the present paper.

There is interest from the superconductor-device community, particularly regarding whether the Higgs plays any role in the imaginary part of $\sigma(\omega)$, which persists to low frequencies. This contribution leads to the well known $1/\omega$ term that reflects the superfluid density. The two are connected by gauge invariance and the associated sum rule.
Through gauge invariance, our Fig.~\ref{fig3} emphasizes how the Higgs mode must necessarily lead to a negative weight in the superfluid density and thus show up in the anisotropic contributions to the imaginary conductivity at low frequencies which are relevant to device applications.
Interestingly, some indications of this effect (which is observed in the THz regime)
appear to be present in experiments~\cite{privatecommunication}.

{\it Acknowledgment.} We thank Andrew Higginbotham for very helpful discussions
and Ryo Shimano for his communications.
We also acknowledge the University of Chicago's Research Computing Center for their support of this work. R.~B. was supported by the Department of Physics and Astronomy, Dartmouth College.

\bibliography{ref2}

\begin{thebibliography}{41}
\bibitem [{foo()}]{footnote}%
  \BibitemOpen
  \href@noop {} {\bibinfo  {journal} {The Higgs mode and the amplitude mode are identical only in a charged theory}\ }\BibitemShut {NoStop}%
\end{thebibliography}

\end{document}


\title{
       Supplemental Material for  
      The Higgs- Amplitude mode in the optical conductivity in the presence of a
supercurrent: Gauge invariant forumulation}

\setcounter{equation}{0}
\setcounter{figure}{0}
\setcounter{table}{0}
\setcounter{page}{1}
\makeatletter
\renewcommand{\thesection}{S\arabic{section}}
\renewcommand{\theequation}{S\arabic{equation}}
\renewcommand{\thefigure}{S\arabic{figure}}
\renewcommand{\bibnumfmt}[1]{[S#1]}

 \author{Ke Wang}
\affiliation{Department of Physics and James Franck Institute, University of Chicago, Chicago, Illinois 60637, USA}
\affiliation{Kadanoff Center for Theoretical Physics, University of Chicago, Chicago, Illinois 60637, USA}

\author{Rufus Boyack }
\affiliation{Department of Physics and Astronomy, Dartmouth College, Hanover, New Hampshire 03755, USA}

     \author{K. Levin}
     \affiliation{Department of Physics and James Franck Institute, University of Chicago, Chicago, Illinois 60637, USA}
     \maketitle

\section{Characterizing the Higgs}

\begin{figure}
    \centering
    \includegraphics[width=3.0in,clip]{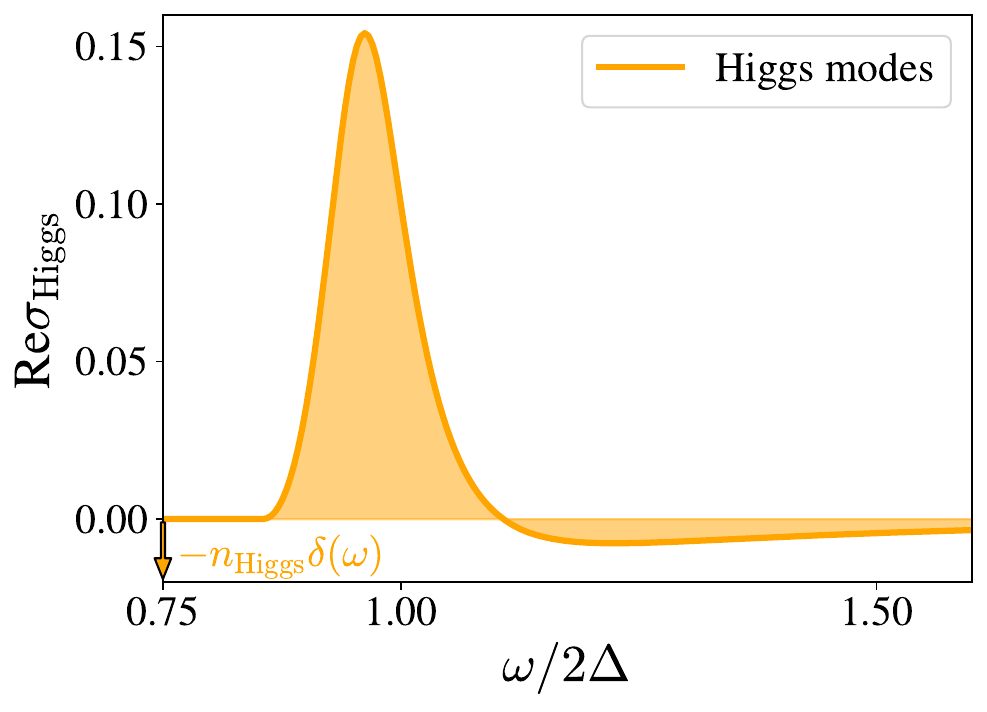}
    \caption{ (Color online). Plot of $\text{Re}\sigma_{\text{Higgs}}(\omega)$ in units of $\sigma_0 = e^2 m k_F / (2\pi^2 \hbar)$.
What is plotted here is
the projection of the conductivity tensor along the $\hat{Q}$-direction We take as typical parameters $\Delta \tau = 0.2$ and $\Delta / E_F = 0.2$. The Higgs component of the conductivity is characterized by two key features: a positive peak around $2\Delta$ and a negative superfluid weight at $\omega = 0$.
    }
    \label{fig1a}
\end{figure}

Fig.~\ref{fig1a} plots the isolated Higgs contribution. One can see that this is consistent with the
f-sum rule, and that it has a negative weight contribution to the conductivity.  There is, as well
a negative delta function component representing the depression in the superfluid density associated
with the Higgs.

\begin{figure}
    \centering
    \includegraphics[width=3.0in,clip]{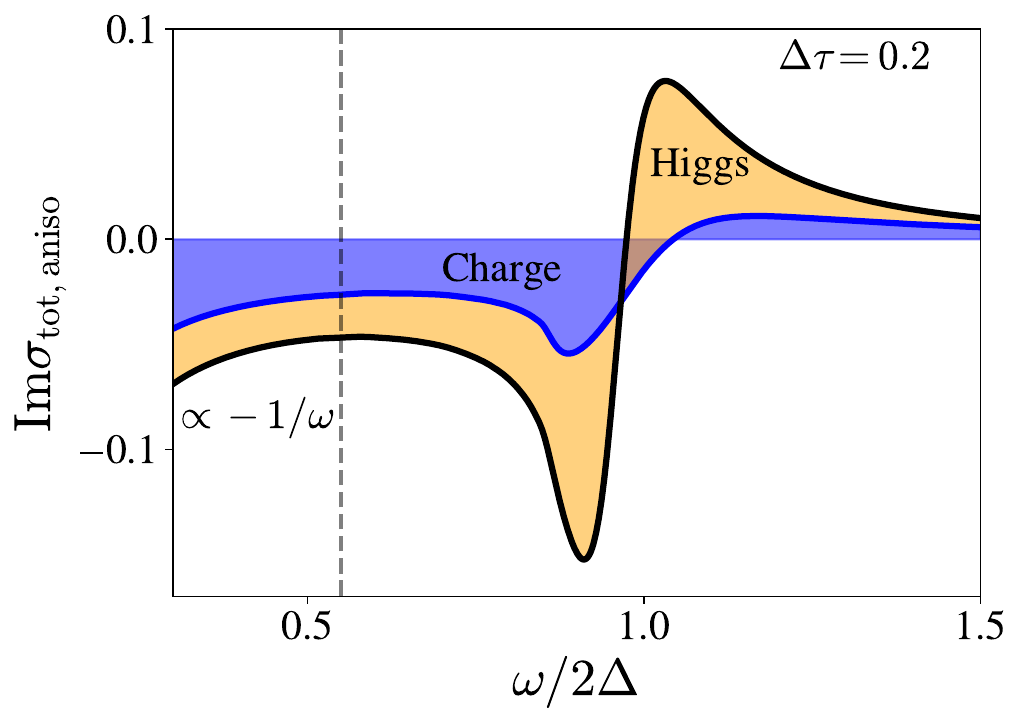}
    \caption{ (Color online). Plot of the imaginary part of $\sigma_{\text{tot},\text{aniso}}$ in units of $\sigma_0$. An asymmetric feature appears around $\omega = 2\Delta$, with the charge component being more asymmetric than the Higgs component. For $\omega < \Delta$, a negative asymptote proportional to $-1/\omega$ emerges, indicating a negative superfluid weight in this anisotropic component.
    }
    \label{fig3a}
\end{figure}

Fig.~\ref{fig3a}  
plots the imaginary
part of the conductivity for moderate disorder which has an interesting feature
directly reflecting the negative superfluid density
contributions discussed
in maintext.
For $\omega > \Delta$ (experimentally accessible regime), the theoretical results align well with the measurements, although this system is noticeably less disordered than in experiment. Around $\omega \sim 2\Delta$, $\sigma_{\text{charge}}$ exhibits greater asymmetry than $\sigma_{\text{Higgs}}$. This reflects the fact that
the overall asymmetry of $\sigma$ is due to $\sigma_{\text{charge}}$
rather than the Higgs contribution. At lower frequencies $\omega < \Delta$, we observe that the curve begins to reflect
the negative superfluid contibution $\propto -1/\omega$.

Disorder plays an important role in the Higgs contributions. The Higgs propagator is defined as $G_H(k) = (S_{11}(k) + 2/g)^{-1}$.
We now examine its fundamental properties. In the clean limit, it can be shown that $G_H(\omega, {\bf k}=0) \propto 1/\sqrt{4\Delta^2 - \omega^2}$, which exhibits a branch-cut singularity, indicating that the Higgs mode is undamped. When a disorder potential is introduced, damping effects arise, and this can be parameterized using
a phenomenological form:
\begin{eqnarray}
\label{S1}
 G_H^{-1}(\omega) = \sqrt{4\Delta^2 - (\omega + i\lambda)^2}.
\end{eqnarray}
Here, $\lambda$ denotes the damping rate of the Higgs mode which we will address more
microscopically. In the two asymptotic limits ($\tau \Delta \ll 1$ and $\tau \Delta \gg 1$), we confirm the form of Eq.~\ref{S1} and identify the scaling properties: $G_H^{-1}(2\Delta) \propto \sqrt{Q^2 / \tau}$ for $\tau \Delta \gg 1$ and $G_H^{-1}(2\Delta) \propto \sqrt{k_F^2 \Delta Q^2 \tau}$ for $\tau \Delta \ll 1$. Here the coefficients
of proportionality are complex. Below we will use these properties to address interesting properties of the calculated conductivity.
Both the charge and Higgs components of the conductivity contain anisotropic (${\bf Q}$-dependent) tensor contributions. These components contribute distinctly, and the value of $\tau$ significantly affects the relative strength of their contributions.

The disorder dependent phenomena depicted in Fig.~2 of the maintext can be explained as follows. $\text{Re}\sigma_{\text{charge}}$ captures the excitations of fermionic pairs. In the presence of disorder, the fermion lifetime scales as $\sim \tau$, and the two-particle propagator behaves as $G^{(2)}(2\Delta) \sim \tau$. By contrast, the Higgs mode exhibits different behavior: in cleaner samples, $G_H(2\Delta) \sim \sqrt{\tau}$. Consequently, when $\tau \Delta \gg 1$, the resonant peak from the two-particle contribution naturally dominates over the Higgs peak. However, in the dirty limit, the peak of $G^{(2)}(2\Delta) \sim \tau$ becomes significantly weaker, while $G_H(2\Delta) \sim 1/\sqrt{\tau}$ forms a sharp peak. Thus, when $\tau \Delta \ll 1$, the Higgs mode contribution to the peak becomes dominant over the charge component. It is important to note, though, that the tail of $\sigma$ is primarily due to $\sigma_{\text{charge}}$, as small $\tau$ causes $\sigma_{\text{charge}}$ to be more evenly distributed across all frequencies.

\section{Damping of the Higgs mode in the dirty limit} \label{ASecII}
The evaluation of the damping rate of the Higgs mode reduces to computing \( G_H^{-1} = S_{11} + 2/g \). We begin with \( S_{ij} \) with Matsubara frequency and zero momentum:
\begin{eqnarray}
S(i\omega_0) = 2\pi T \sum_{ik_0} \left(1 - \frac{1}{2\pi \tau} M(i\omega_0, i\omega_0 - ik_0) E^{33} \right)^{-1} M(i\omega_0, i\omega_0 - ik_0).
\end{eqnarray}
The matrix \( M \) is given by \( M_u = \pi (f_1 \tau_3 + i f_2 \tau_1 + f_3) \), where \( f_\sigma \) is defined by
\begin{eqnarray}
&& f_1 = \left\langle \frac{1}{\tilde{r} + \tilde{r}'} \right\rangle_x, \quad f_2 = -\left\langle \frac{\tilde{\Delta}' \tilde{z} + \tilde{\Delta} \tilde{z}'}{(\tilde{r} + \tilde{r}') \tilde{r} \tilde{r}'} \right\rangle_x, \quad f_3 = \left\langle \frac{\tilde{\Delta} \tilde{\Delta}' \mp \tilde{z} \tilde{z}'}{(\tilde{r} + \tilde{r}') r'} \right\rangle_x.
\end{eqnarray}
Here, \( \tilde{z} = \tilde{\omega}_0 + isx \), and \( \left\langle ... \right\rangle_x \) denotes \( \int_{-1}^{+1} ... dx/2 \), with \( s = v_F Q \) and \( \tilde{r} = \sqrt{\tilde{z}^2 + \tilde{\Delta}^2} \). Note that \( \tilde{\omega}_0 \) and \( \tilde{\Delta} \) need to be determined self-consistently from
\begin{equation}
\label{self}
   \tilde{\omega}_0-\omega_0=\frac{t_+-t_-}{d}  ,  \, \tilde{\Delta}-\Delta=  \frac{\Delta}{d } \log \left( \frac{t_++\tilde{\omega}_+}{t_-+\tilde{\omega}_-}
\right).
\end{equation}
Here $\tilde{\omega}_\pm=\tilde{\omega}_0\pm ik_F Q$, $t_\pm =\sqrt{\tilde{\omega}^2_\pm+{\tilde \Delta}^2}$, and $d=4 i k_F Q \tau$, where $\tau$
is the impurity scattering time. Similarly, \( \tilde{\omega}' \) and \( \tilde{\Delta}' \) are derived from \( \omega_0 - k_0 \). Consequently, the expression for \( S_{11} \) simplifies to
\begin{eqnarray}
\label{S11}
&& S_{11} = 2\pi T \sum_{p_0} \frac{f_3 - \left[f_1 + D/2\tau \right]}{1 - f_1/\tau - D/4\tau^2}, \quad D = f_2^2 + f_3^2 - f_1^2.
\end{eqnarray}
Here, \( D = 0 \) if \( Q = 0 \). The evaluation of \( f_\sigma \) involves an angular average over \( x \). In the \( \tau \Delta < 1 \) limit, \( f_\sigma \) is of the order \( \tau \), allowing the use of the integrand function at \( x = 0 \) as an approximation for \( f_\sigma \). The deviation from the \( x = 0 \) value is minor, on the order of \( O(v_F^2 Q^2 \tau^3) \). Thus, in the dirty limit, we approximate
\begin{eqnarray}
\label{S12}
G^{-1}_H &=& \int_C \frac{d\omega_0}{e^{i\beta \omega_0} + 1} \left[ \frac{(\tilde{\Delta} \tilde{\Delta}' - \tilde{\omega}_0 \tilde{\omega}_0')/\tilde{r} \tilde{r}' - 1}{\tilde{r} + \tilde{r}' - 1/\tau} + \frac{1}{r} \right].
\end{eqnarray}
Here, \( C \) represents a contour encircling the real axis, \( r = \sqrt{\omega_0^2 + \Delta^2} \), and \( \tilde{r} \) is evaluated at \( x = 0 \) in this integral, namely $\sqrt{\tilde{\omega}_0^2 + \tilde{\Delta}}$. In the dirty limit, we find:
\begin{eqnarray}
\label{S13}
\sqrt{\tilde{\omega}_0^2 + \tilde{\Delta}} \simeq \sqrt{\omega_0^2 + \Delta^2} + \frac{1}{2\tau} + \frac{k_F^2 Q^2 \tau \Delta^2}{3 (\Delta^2 + \omega_0^2)}.
\end{eqnarray}
Note that \( \sqrt{\tilde{\omega}_0^2 + \tilde{\Delta}} \rightarrow \sqrt{\omega_0^2 + \Delta^2} \eta \) as \( \tau \rightarrow 0 \), with \( \eta = 1 + 1/2\tau \sqrt{\omega_0^2 + \Delta^2} \), typically observed in dirty BCS systems. Using Eq.~\ref{S13}, we solve
\begin{eqnarray}
\tilde{\omega}_0 = \omega_0 \eta + \frac{1}{3} \frac{\omega_0 \Delta^2}{(\Delta^2 + \omega_0^2)^2} k_F^2 Q^2 + O(\tau), \quad
\tilde{\Delta} = \Delta \eta - \frac{1}{3} \frac{\omega_0^2 \Delta}{(\Delta^2 + \omega_0^2)^2} k_F^2 Q^2 + O(\tau). \label{S14}
\end{eqnarray}
Interestingly, \( \tilde{\omega}_0 \) deviates from the dirty BCS value by a constant term proportional to \( Q^2 \) even as \( \tau \rightarrow 0 \).

At zero temperature, we replace the contour integral with a real-axis integral and expand based on Eqs.~\ref{S13} and ~\ref{S14}:
\begin{eqnarray}
&& \tilde{r} + \tilde{r}' - 1/\tau = r + r' + \frac{k_F^2 Q^2 \tau \Delta^2}{3} \left(\frac{1}{r^2} + \frac{1}{r'^2}\right), \\
&& \frac{\tilde{\Delta} \tilde{\Delta}' - \tilde{\omega}_0 \tilde{\omega}_0'}{\tilde{r} \tilde{r}'} \simeq \frac{\Delta \Delta - \omega_0 \omega_0'}{r r'} + 2\tau k_F^2 Q^2 \frac{(\Delta_0^2 + \omega_0 \omega_0')}{3 rr'} \left(\frac{\Delta \omega_0}{r^3} + \frac{\Delta \omega'_0}{r'^3}\right).
\end{eqnarray}
Consequently, performing the integral in Eq.~\ref{S12} yields:
\begin{eqnarray}
\label{S17}
G^{-1}_B(i\omega_0) \simeq \sqrt{\omega_0^2 + 4\Delta^2} + 2\frac{k_F^2 q^2 \tau \Delta^2_0}{\sqrt{4\Delta_0^2 + k_0^2}} b(\omega_0).
\end{eqnarray}
Here, \( b(\omega_0) \) represents a complex function obtained from the integral. The function \( b(\omega_0) \) is analytic in the complex plane, except for branch cuts along the imaginary axis, where its values are bounded by \( |\omega_0| \). One can rewrite Eq.~\ref{S17} in a "non-perturbative" form:
\begin{eqnarray}
G^{-1}_B(i\omega_0) \simeq \sqrt{k_0^2 + 4\Delta^2 + k_F^2 q^2 \tau \Delta^2_0 b(\omega_0)}.
\end{eqnarray}
Finally, we perform the analytical continuation $i\omega_0\rightarrow \Omega+i\delta$ with $\delta>0$:
\begin{equation}
G^{-1}_B(\Omega + i\delta, 0) \simeq \sqrt{4\Delta^2 - \Omega^2 + k_F^2 Q^2 \tau b(-i\Omega + \delta)}.
\end{equation}
Thus, we confirm the form of Eq.~\ref{S1} within the dirty limit, indicating that the asymptotic scaling of the damping rate \( \lambda \) at \( \Omega = 2\Delta \) is proportional to \( k_F^2 Q^2 \tau \).

\section{Impurity effects on vertex operators}
Vertex operators are renormalized by the disorder potential, as we pointed out in Eq.~17 of main-text. Here we provide details the renormalization of superconductor pairing field and
current operators.

Firstly, we consider how the disorder potential renormalizes the superconducting pairing field. The bare vertex operators are given by \(\gamma = \tau_i\), with \(i = 0, 1, 2, 3\). The renormalized \(\Gamma\) can be expanded as $ \Gamma_i = \sum_{j=0}^3 T_{ij} \tau_j.
$
Here, \(T\) is found to be \(T = \left( 1 - (2\pi \tau)^{-1} M E \right)^{-1}\), where \(E = (\tau_3, 0; 0, -\tau_3)\), and each matrix element of \(M\) is determined by
\[
M_{lm} = \frac{\text{tr}}{2} \int d\xi_{\bf k} \frac{d\Omega}{4\pi} G(k) \tau_l G(k-q) \tau_m,
\]
where \(q = p_+ - p_-\) and \(d\Omega = \sin\theta \, d\theta \, d\varphi\). Therefore, one can determine \(\Gamma_i\) from \(M\) and \(T\).

Then we consider the current operator. The bare current operator for a superconductor carrying a uniform supercurrent is
\begin{eqnarray}
j({\bf p})=\frac{{\bf p}\tau_0+{\bf Q} \tau_z}{m},\quad \hat{j}({\bf k})=\sum_{\bf p } \Psi^\dagger({\bf p}-{\bf k}/2) j({\bf p}) \Psi({\bf p}+{\bf k}/2).
\end{eqnarray}
The effects of disorder renormalize the current operator as follows:
\begin{eqnarray}
 \Gamma_{\text{J}}(\mathbf{p})=j({\bf p})+n_i \sum_{\mathbf{p}'} |V_n(\mathbf{p}-\mathbf{p}')|^2 \tau_3 G_{\mathbf{p}' }(\omega)
  \Gamma_{\text{J}}(\mathbf{p}')
    G_{\mathbf{p}' }(\omega-\Omega)\tau_3.
\end{eqnarray}
Here, \( V \) represents the impurity potential. We consider non-magnetic impurities as \( V({\bf r})\tau_z \) and assume a contact interaction for this disorder.  An important observation is that the vertex can always be split into two parts:
\begin{eqnarray}
\Gamma_{\text{J}}(\mathbf{p})=\frac{{\bf p}\Lambda_0+{\bf Q} \Lambda_1}{m}.
\end{eqnarray}
Here, \( \Lambda_0 \) and \( \Lambda_1 \) are two \( 2\times 2 \) matrices. For a contact potential, one can prove \( \Lambda_0=\tau_0 \). This follows because the integration over \( G\Gamma_{\text{J}} G \) in \( p' \) only provides corrections along the \( {\bf Q} \)-direction. Thus, we have \( \Lambda_0=\tau_0 \). The matrix \( \Lambda_1 \), associated with \( {\bf Q} \), satisfies the equation:
\begin{eqnarray}
{\bf Q}\Lambda_1&=&{\bf Q}\tau_z+\frac{1}{2\pi \tau} \int \frac{dx}{2} \int d\xi_{p'} \tau_3 G_{\mathbf{p}' }(\omega)
  \left({\bf p}'\tau_0+{\bf Q} \Lambda_1 \right)
    G_{\mathbf{p}' }(\omega-\Omega)\tau_3 . \label{S4}
\end{eqnarray}

Projecting Eq.~\ref{S4} along \( {\bf Q} \) it follows that
\begin{eqnarray}
\Lambda_1
    &=&\tau_z+ \frac{1}{2\pi \tau} \frac{p_F}{Q} \int d\xi_p \frac{dx}{2} \tau_3 G_{\mathbf{p}' }(\omega)
  x \tau_0
    G_{\mathbf{p}' }(\omega-\Omega)\tau_3+  \frac{1}{2\pi\tau} \int d\xi_p  \frac{dx}{2} \tau_3 G_{\mathbf{p}' }(\omega)   \Lambda_1 G_{\mathbf{p}' }(\omega-\Omega)\tau_3  .\label{S5}
\end{eqnarray}
Note that the second term does not include \( \Lambda_1 \). We define \( c_0, c_1, \) and \( \Pi_l \) as follows:
\begin{eqnarray}
c_0+c_1 \tau_1\equiv \frac{1}{2\pi\tau} \frac{p_F}{Q}
\int_{-1}^1 \frac{dx}{2} x\tau_3 \Pi_0 \tau_3,  \quad \text{and} \, \quad \Pi_l&\equiv& \int d\xi_p    G_{\mathbf{p}' }(\omega)   \tau_l G_{\mathbf{p}' }(\omega-\Omega).
\end{eqnarray}
Thus, Eq.~\ref{S5} can be rewritten as
\begin{eqnarray}
\Lambda_1
    &=&c_0+c_1\tau_x+\tau_z +  \frac{1}{2\pi\tau} \int d\xi_p  \frac{dx}{2} \tau_3 G_{\mathbf{p}' }(\omega)   \Lambda_1 G_{\mathbf{p}' }(\omega-\Omega)\tau_3.
\end{eqnarray}
Expanding \( \Lambda_1 \) in Pauli matrices, it follows that $
\Lambda_1=\sum_i B_i \tau_i,
$
and we find \( {\vec B}={\vec c}\left(1-\frac{1}{2\pi \tau}    M   E^{33} \right)^{-1}. \)
Here we define vectors \( {\vec B}=(B_0,B_1,B_2,B_3),{\vec c}=(c_0,c_1,0,1) \) while \( M \) and \( E^{33} \) are specified in Eq.~21 of the main text.  Finally, the renormalized current vertex within the supercurrent framework is
\begin{eqnarray}
\Gamma_{\text{J}}({\bf p};\omega,\omega')=\frac{{\bf p}\tau_0 }{m}+\frac{{\bf Q}\tau_i }{m} \left[{\vec c}\left(1-\frac{1}{2\pi \tau} \langle M \rangle E^{33} \right)^{-1}\right]_i.
\end{eqnarray}